\documentclass[aps,twocolumn,pra,showpacs,superscriptaddress,floatfix]{revtex4-2}

\usepackage{latexsym,amsmath,amssymb,amsfonts,graphicx,color,amsthm}
\usepackage{enumerate}
\usepackage{braket}
\usepackage{adjustbox}
\usepackage{footnote}
\usepackage[dvipsnames]{xcolor}
\usepackage{tipa}
\usepackage{textcomp}
\usepackage{fontenc}
\usepackage{mathrsfs}
\usepackage{colortbl}
\usepackage{txfonts}
\usepackage{graphics,graphicx}
\usepackage{mathtools}
\usepackage{epsfig}
\usepackage{epstopdf}
\usepackage{dsfont}
\usepackage[colorlinks=true,linkcolor=blue,urlcolor=magenta,citecolor=red]{hyperref}
\usepackage[normalem]{ulem}
\usepackage[caption=false,labelformat=simple]{subfig}

\begin{document}


\title{Multi-bit quantum random number generator from path-entangled single photons}

\author{K. Muhammed Shafi}
\affiliation{Quantum Optics \& Quantum Information, Department of Instrumentation and Applied Physics,\\ Indian Institute of Science, Bengaluru 560012, India}
\author{Prateek Chawla}
\affiliation{The Institute of Mathematical Sciences, C. I. T. Campus, Taramani, Chennai 600113, India}
\affiliation{Homi Bhabha National Institute, Training School Complex, Anushakti Nagar, Mumbai 400094, India}
\author{Abhaya S. Hegde}
\affiliation{Quantum Optics \& Quantum Information, Department of Instrumentation and Applied Physics,\\ Indian Institute of Science, Bengaluru 560012, India}
\author{R. S. Gayatri}
\affiliation{Quantum Optics \& Quantum Information, Department of Instrumentation and Applied Physics,\\ Indian Institute of Science, Bengaluru 560012, India}
\author{A. Padhye}
\affiliation{Quantum Optics \& Quantum Information, Department of Instrumentation and Applied Physics,\\ Indian Institute of Science, Bengaluru 560012, India}
\author{C. M. Chandrashekar}
\email{chandracm@iisc.ac.in}
\affiliation{Quantum Optics \& Quantum Information, Department of Instrumentation and Applied Physics,\\ Indian Institute of Science, Bengaluru 560012, India}
\affiliation{The Institute of Mathematical Sciences, C. I. T. Campus, Taramani, Chennai 600113, India}
\affiliation{Homi Bhabha National Institute, Training School Complex, Anushakti Nagar, Mumbai 400094, India}


\begin{abstract}
Measurement outcomes on quantum systems exhibit inherent randomness and are fundamentally nondeterministic. This has enabled quantum physics to set new standards for the generation of true randomness with significant applications in the fields of cryptography, statistical simulations, and modeling of the nondeterministic behavior in various other fields. In this work, we present a scheme for the generation of multi-bit random numbers using path-entangled single photons. For the experimental demonstration, we generate a path-entangled state using single photons from spontaneous parametric down-conversion (SPDC) and assign a multi-qubit state for them in path basis.  One-bit and two-bit random numbers are then generated by measuring entangled states in the path basis. In addition to passing the NIST tests for randomness, we also demonstrate the certification of quantumness and self-certification of quantum random number generator (QRNG) using Clauser, Horne, Shimony and Holt (CHSH) inequality violation. We also record the significantly low autocorrelation coefficient from the raw bits generated and this along with CHSH violation rules out multi-photon events and ensure the protection from photon splitting attack. Distribution of photons along multiple paths resulting in multiple bits from one photon extends the limit on bit generation rate imposed by the detection dead time of the individual detector. Thus, the path-entangled states can generate higher bitrates compared to scheme using entangled photon pair which are limited by the coincidence counts. We demonstrate this by generating a high rate of about 80 Mbps when the single photon detector saturates at around 28 Mcps and still show violation of CHSH inequality.
\end{abstract}



\maketitle

\section{Introduction}

True random numbers are a valuable resource with applications in cryptography\,\cite{Sha49, Bla14} and cybersecurity, where security is assured by unpredictability\,\cite{Eke91, BBB92, GRT02}.  Randomness in general is a very useful resource\,\cite{Hay01} in various science and engineering applications like statistical simulations\,\cite{MU49, RK16} and modeling of nondeterministic behavior across fields like artificial intelligence, neural networks, and genetic algorithms. There are many statistical tests which can certify the randomness of the observed sequence\,\cite{Mau92, Mar96, Kol98, RSN01}. However, even if a sequence of random numbers passes all the statistical tests, it is almost impossible to discriminate between a predetermined random string of bits that comes from a dishonest provider or malicious random number generator (RNG) and a truly random sequence. Genuine randomness cannot be generated and unconditionally certified using purely classical methods. In other words, true randomness can only occur through physical processes involving inherent randomness. Till date, the only systems displaying processes with intrinsic randomness are those based on the principles of quantum physics\,\cite{Sch70}. Recent developments in controlling quantum systems and extracting intrinsic randomness from measurement outcomes have led to new standards in generation and certification of true randomness for use in practical applications\,\cite{CE17, JJN21}. The use of nonlocal correlation between two particles has been shown to reduce the device dependence of quantum random number generators (QRNGs)\,\cite{BHK05, MAG06, AM16}. Device-independent QRNGs can not only allow one to validate randomness, but also to certify that it is private randomness; the produced random numbers are ensured to be unknown to an adversary. Therefore, nonlocality of pure entangled states has been an invaluable resource for information processing tasks such as random number generation\,\cite{PAM10, MYC16}, randomness expansion\,\cite{CK11, CY14}, quantum key distribution\,\cite{BW92, AGM06, PAB09} and privacy amplification protocols\,\cite{Col09, CR12}.

Various approaches to build an efficient QRNG have been developed and evolved over the years as it remains a topic of active interest. For example, QRNG based on quantum shot noise limit\,\cite{SAL11}, phase noise\,\cite{QCL10}, quantum vacuum fluctuations\,\cite{HPL19, ZZH19, DOH21, FSP21}, phase fluctuations\,\cite{XQM12},  single photon emitters in gallium nitride\,\cite{LCF20}, defect centers arising from nitrogen vacancies in diamond\,\cite{CGW19}, hexagonal boron nitride single photon emitters\,\cite{WKT20} along with several other approaches have been reported\,\cite{DYS08, JAW00, WK10, JJD20, LHC19, GLK21, LLW21}. Efforts have also gone into the generation of truly random  bits from the entangled pairs of photons obtained from spontaneous parametric down-conversion (SPDC) processes\,\cite{QMD04, BB10, XSW16, KCK09} and from single-photon entanglement states\,\cite{MLA21, LAM22}. Each of these protocols involves sources that can be broadly classified into three categories, namely, trusted device, device-independent sources\,\cite{AMV18}  and semi-device independent sources\,\cite{CE17}. Although device-independent QRNGs based on nonlocality and entangled states are more secure compared to the sources from the other two categories, they are considered unsuitable for applications which require higher rates of generation of random numbers. The coincidence counts in the measurements of two entangled photons limits the bitrates of these sources.  Any proposal in the direction of increasing the generation rate of random numbers based on violation of the Clauser, Horne, Shimony, and Holt (CHSH) inequality will thus increase the scope of device-independent QRNGs. 

In this work, we present a scheme for the generation of multi-bit random numbers using path-entangled single photons. Using different configurations of path-entangled single photon states\,\cite{PLM20, SGP21}, different distributions along the path can be engineered to extract random numbers. The violation of CHSH inequality ensures the intrinsic randomness in the scheme despite having control over the components in the device, thereby making it a device-independent and self-testing QRNG. For the experimental demonstration, we have used single photons from SPDC process resulting from a nonlinear periodically-poled potassium titanyl phosphate (PPKTP) crystal. We demonstrate the generation of one-bit and two-bit random numbers from each detected single photon in position basis by assigning two-qubit state to the position basis of the path-entangled photons.  The raw bits generated as such are able to pass most of the 15 standard NIST tests used for certifying randomness. After a round of XOR or Toeplitz post-processing~\cite{LEV87, Kra94}, the generated bits are capable of passing all the randomness tests in the NIST suite~\cite{RSN01}. We also demonstrate the certification of quantumness using CHSH inequality violation.  Significantly low auto correlation coefficient of the order of $10^{-3}- 10^{-4}$ recorded from the raw bits rules out multi-photon events and protects QRNG from photon splitting attack.
We also demonstrate the transition of from CHSH inequality violation and to validity with decrease in entanglement visibility and the same results in increase of autocorrelation coefficient. 
The path entanglement-based scheme demonstrated in this work presents major advantages over current implementations, including, but not limited to: 
\begin{enumerate}
\item Each detected photon translates to a two-bit random number. Moreover, the approach presents scalability to generate a higher number of bits either by increasing the number of paths and/or by including other degrees of freedom such as polarization.
\item Compared to other two-particle entanglement-based schemes that use rarer coincidence counts, this scheme uses single photon detection events to generate random bits. Thus, the scheme benefits from the improvements in bitrates.
\item Due to path entanglement of states, photon detection events gets distributed among multiple detectors, allowing for a higher output bitrates as the limiting factor of detector dead time is effectively bypassed. We are able to achieve an output bitrate of about $45 \times 10^6$ bps (bits per second) and  $80 \times 10^6$ bps with path entanglement certification of heralded and unheralded single photons, respectively. Such high rate is recorded despite each detector is being limited to a photon count of $28 \times 10^6$ cps (counts per second) on its own.  It is possible to achieve bitrates of up to 100--130 Mbps by increasing pump power in our setup before the detectors begin to saturate and multi-photon detection dominates resulting in absence of CHSH violation.
\end{enumerate}

This paper is organized as follows.  In section \ref{theory}  we have outlined the theoretical description of the path-entangled photon state and the scheme for generating multi-bit random numbers, along with discussion on the visibility of path-entangled state and post-processing analysis. In section \ref{Expt} experimental setup to generate one-bit and two-bit random numbers is described. Experimental results, tests of QRNG and details of certification are presented in section \ref{Expt}. We conclude and summarize our results in section \ref{conc}.


\section{Multi-bit QRNG from path-entangled state}
\label{theory}

Measurements on quantum states of light with inherent randomness in various parameters is one of the most commonly employed methods to extract quantum randomness. Quantum states in an entangled basis have been used in device-independent schemes for QRNG. Various choices of apparatus exist for the generation of quantum states of light, however, the affordability of single-photon sources and the ready availability of detectors have made them a convenient choice for constructing QRNGs.  

An elementary realization of a QRNG is based on excessively attenuated laser sources. Single photons generated via attenuation techniques is only an approximation to single photons and are thus affected by photon statistics of coherent states, along with single photons many two and multi-photons states will also occur. Single photon events are isolated by discarding all coincidence detection when they are passed through the beam splitter. Fidelity of single photons from attenuated source is limited due to significant level of post selection~\cite{BT56}.  A mature and convenient method to enter the single photon regime is to employ SPDC of photons caused by a $\chi^{2}$ non-linear crystal. By heralding one of the photons from the down converted pair, single photons are generated.  The pair of photons resulting from the down-conversion can also be entangled in polarization degree of freedom with a careful post selection of the down converted photons pairs.  Among various schemes for QRNG using single and entangled photons,  device-independent schemes have been reported using entangled photon sources.  However, a QRNG using entangled photon pairs relying on coincidence events suffers from low bit generation rates. Here we present a scheme using single path-entangled photons certified via CHSH inequality violation\,\cite{PLM20, SGP21} to convert every detected photon into a random multi-bit sequence. 

A single photon state that is in superposition along $m$ path can be written in the form,
\begin{align}
\label{path}
|\Psi\rangle_{P}  = \sum_{x=1}^{m} \alpha_{x} |x\rangle.
\end{align}
In path-entangled representation the state exists in a Hilbert space $\mathcal{H} = \bigotimes_{i=1}^{m} \mathcal{H}_{p_i}$, where $\mathcal{H}_{p_i}$ is the Hilbert space of the $i^{th}$ path. Each $\mathcal{H}_{p_i}$ will be spanned by  $\{ \ket{0}_{p_i}, \ket{1}_{p_i} \}$ representing the absence and presence of the photon in the path, respectively,
\begin{align}
\label{path-ent}
|\Psi\rangle_{PE}  = \sum_{x=1}^{m} \alpha_{x} |x\rangle =  \sum_{p_{1},...,p_{m} \in \{0, 1\}} \alpha_{p_{1}, ..., p_{m}} \ket{p_{1}\,...\,p_{m}}.
\end{align}
The output state of the photon passing through a 50:50 beam splitter in path-entangled representation will be,
\begin{align}\label{eq:path-photonBS}
|\Psi\rangle_{PE}  =  \frac{1}{\sqrt 2}\Big [ |0\rangle_{P_1}|1\rangle_{P_2}  + i |1\rangle_{P_1}|0\rangle_{P_2}  \Big ]. 
\end{align}
The CHSH parameter calculated for the above state following the procedure given in\,\cite{SGP21} is $S = 2\sqrt{2}$. This path-entangled state which violates CHSH inequality can  be further split into $m$ level using array of beam splitters. By assigning a multi-bit state to each path, it can be rewritten as an entangled multi-qubit state. When $m=2^n$, the multi-qubit representation will be, 
\begin{align}\label{eq:photon-BS2}
\ket{\Psi}_{MQ}  =  \sum_{q_{1},...,q_{n} \in \{0, 1\}} \alpha_{q_{1},  ..., q_{n}} \ket{q_{1}\,...\,q_{n}}.
\end{align}
The measurement on the preceding path-entangled state collapses the state to one of the $m$ paths and returns a multi-bit value assigned to that path. With some redundancy, a $n-$qubit state can be used to represent $m$ paths even when $m < 2^n$. 
\begin{figure}[ht!]
   \centering
    \includegraphics[width=0.47\textwidth]{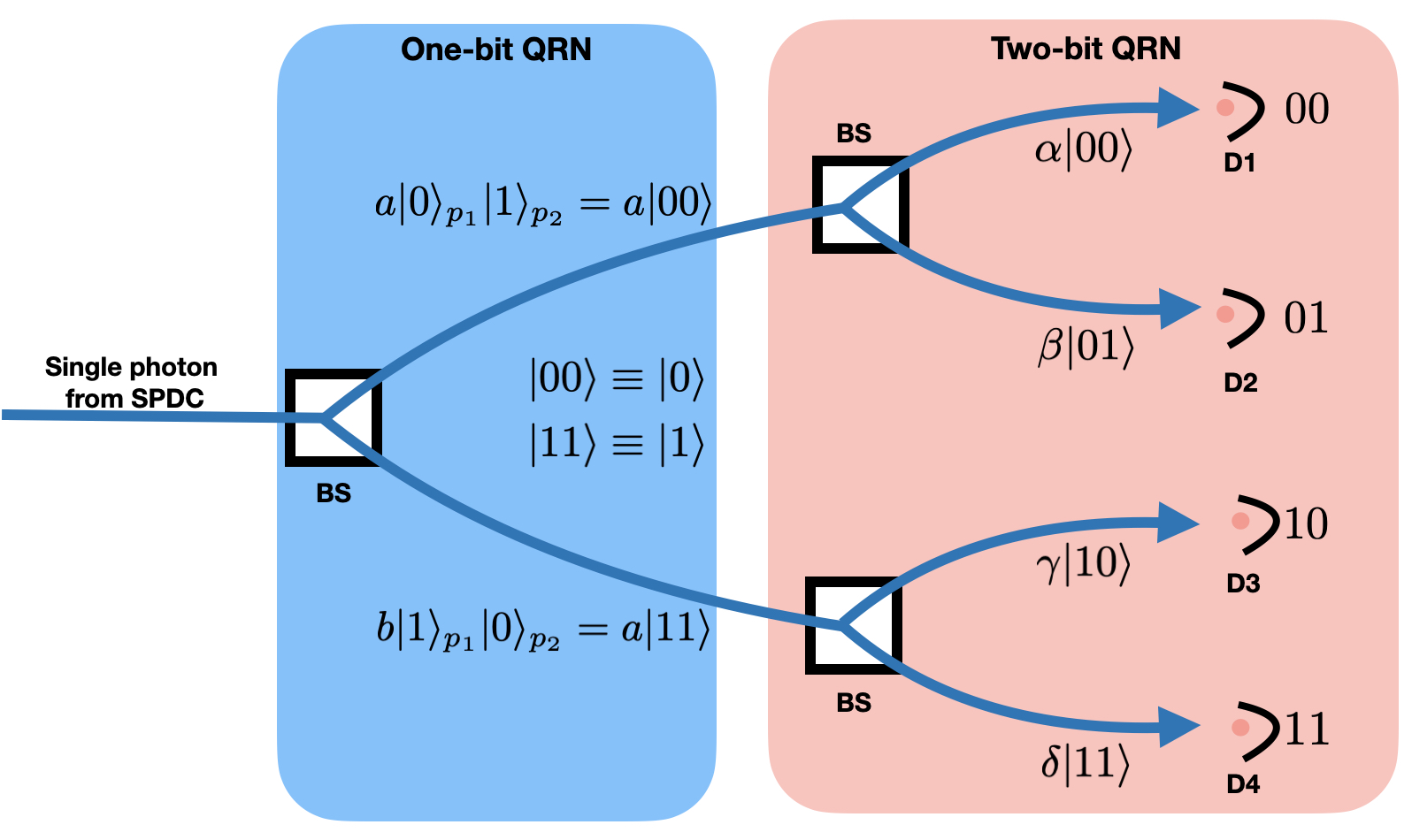}
    \caption{Schematic representation of a scheme to generate path-entangled single photon states mapped to a two-qubit state for each photon detection. A measurement operation after passing the beam through the first beam splitter will return one-bit random numbers, and after passing the second stage beam splitter pair, we can extract two-bit random numbers.}
   \label{fig:Schematic}
\end{figure}
Fig.\ref{fig:Schematic} illustrates the configuration of beam splitters in the path of a single photon to generate a two qubit state. One- and two-bit random numbers can be extracted from the measurement outcomes on this state with the detectors and bit assignment corresponding to the detector. The path-entangled state after passing through the first beam splitter will take the form given in Eq.\,\eqref{eq:path-photonBS}. When the state is further passed through a beam splitter given by the operation, 
\begin{align}
B(\theta) = \begin{bmatrix}
	~\cos(\theta) & ~~i\sin(\theta) \\
	i\sin(\theta) & ~~\cos(\theta)
	\end{bmatrix}
\end{align}	
along each path, the two-qubit state representation will be,
\begin{align}
|\Psi_{2Q}\rangle =  \frac{1}{\sqrt{2}} \Bigg [ & \cos(\theta) |00\rangle + i \sin(\theta) |01\rangle \nonumber \\ & + i \sin(\theta) |10\rangle + \cos(\theta)|11\rangle \Bigg ].
 \end{align}
Measurement will result in two-bit random number.  
If we  consider polarization degree of freedom of photons in addition to path encoding we will obtain,
\begin{align}\label{eq:pol-path-photon}
\begin{split}
\ket{\Psi}_{P-PE}  = \sum_{p_{1},...,p_{m} \in \{0, 1\}} \Big[ &\alpha_{p_{1}, p_{2}, ..., p_{m}} \ket{H}\otimes\ket{p_{1}\,p_{2}\,...\,p_{m}}  \\
& + \beta_{p_{1}, p_{2}, ..., p_{m}} \ket{V}\otimes\ket{p_{1}\,p_{2}\,...\,p_{m}} \Big]
\end{split}
\end{align}
where $\{\ket{H},\ket{V}\}$ are the eigenbasis of the polarization Hilbert space, and $\alpha_k, \beta_k$ are chosen such that the state is normalized. This can be used to generate another random bit in addition to bits from the path-entangled states. One of the established practical ways to generate the above described photon-path entangled state with the multi-qubit mapping is via quantum walks\,\cite{ASN20}. Generation of multi-bit quantum random numbers using discrete-time quantum walk and increase in intrinsic randomness and entropy with increase in path degree of freedom has also been reported using a similar construct\,\cite{SC19}.

Therefore, assuming ideal photon generation and detection, the measurement output from the above scheme is guaranteed to be intrinsically random where quantumness is certified by the violation of CHSH inequality.
The action of a depolarizing noise channel on path-entangled photons reduces entanglement visibility, as indicated by a corresponding decrease in value of the CHSH parameter $S$. Loss of quantumness is then characterized by $S \leq 2$. The visibility of path-entangled quantum state can also be controlled by changing the splitting ratio directly along the paths or by controlling the polarization degree of freedom using a combination of quarter wave plates and half wave plate along with polarizing beam splitter. Thus, by controlling the visibility, the effect of depolarizing noise on path-entangled photons can be mimicked.  When $m=2$, visibility can be obtained from the number of counts of photons detected on detectors $D_{1}$ and $D_{2}$,
\begin{align}
\mathcal{P} = \frac{(C_{D_{1}} + C_{D_{2}} ) -   \lvert C_{D_{1}} - C_{D_{2}} \rvert  }{ (C_{D_{1}} + C_{D_{2}} ) +  \lvert  C_{D_{1}} - C_{D_{2}}\rvert   }.
\end{align}
It has been shown that achieving a value $S > 2$ corresponds to the setup wherein visibility $\mathcal{P}$ exceeds $0.7$\,\cite{SGP21}. This is a useful technique to generate and test the quality of random numbers generated for different visibility values as it determines the threshold to certify the quantumness of the devices.
\subsection*{Post-processing analysis}
Theoretically, a QRNG should be able to produce truly random bitstrings in its output. However, in practical realizations, it is inevitable that some other classical signals (noise) will affect the output correlations. Optical fiber coupling and detector correlations are just two examples of many such possibilities.  While noise cannot be controlled by the device designer, it remains a liability and can affect the randomness of the output. To illustrate the correlation of QRNG to effects that mimic noise, we may consider an example of imperfect detector $D$, which can detect a photon in the optical mode subspace spanned by  $\{ \ket{D_{\uparrow}}, \ket{D_{\downarrow}} \}$, denoting the presence of no photon or a single incoming photon respectively. Assume that the detection happens with the positive operator-valued measure (POVM)  $\{K^\uparrow, K^\downarrow \}$, and that probability of detection is bounded by $\epsilon$. Thus,
\begin{align}
\begin{split}
	K^\downarrow &= \epsilon \ket{D_{\downarrow}}\bra{D_{\downarrow}} \\
	K^\uparrow &= \mathds{1} - K^\downarrow  
\end{split}
\end{align}
where $\mathds{1}$ is the identity operation in the optical mode subspace. To describe $K$ as a POVM, we can consider the extension of $K$ into an additional space of the subsystem $S$, which decides if the detector will detect this incoming photon. {Thus the detector $D$ is described by the subsystem $K$, which detects the incoming photon, and the subsystem $S$, which determines if the incoming photon can be detected at all. In other words, the subsystem $S$ controls the sensitivity of the detector, while the subsystem $K$ performs the actual detection via a POVM.  Let us assume the Hilbert space of system $S$ is spanned } by the basis $\{ \ket{S_{\uparrow}}, \ket{S_{\downarrow}} \}$, denoting the case of no detection and successful detection, respectively. 
Then the outcome of the projective measurement $\{K^\uparrow, K^\downarrow \}$ applied to {the joint system of the detector and an external system $E$} with the following purification 
\begin{equation}
\sigma_{DE} = \sqrt{1-\epsilon}\ket{S_{\uparrow}}\otimes\ket{E_{\uparrow}} + \sqrt{\epsilon} \ket{S_{\downarrow}}\otimes \ket{E_{\downarrow}}
\end{equation}
will result in  information about the sensitivity of the detector. Here, $E$ is chosen to be a purifying system, i.e. the joint state of the QRNG setup and $E$ is a pure state. For simplicity, we model our scheme assuming that the number of photons emitted by the source follows a Poisson-distribution. A photon detection event in a detector is assigned the bit $1$ whereas no detection corresponds to bit $0$. Assuming an imperfect detector in such a case reveals that with a higher number of photons emitted by the source, the guessing probability of the state $11$ tends towards a higher quantity. This results in adversary having information about, say, the detection sensitivity of the subsystem, and thus being able to completely determine the output of the supposed QRNG. Thus, the elementary QRNG model in this regime will output a classical noise and not quantum randomness. In order to guarantee that the output will be free of such correlations, we need to implement post-processing procedures that remove these extraneous correlations and extract the randomness from the generated bit strings.

We use two kinds of post-processing -- one using the XOR extractor~\cite{LEV87}, and another using the Toeplitz extractor~\cite{Kra94}. The XOR extractor is implemented by the XOR operation on bits equidistant from each end of the sequence, and discarding the latter half of the sequence. This method, while useful for data that has a high amount of extractable randomness (min-entropy), begins to fail as the min-entropy decreases. The Toeplitz extractor is able to extract randomness from the output data, however, the losses become higher as the visibility reduces. This is evident from  tests on random bit generated using states with reduced entanglement visibility, as detailed in section \ref{sec:cert}. 

 \noindent
 \begin{figure}
    \includegraphics[width=\columnwidth]{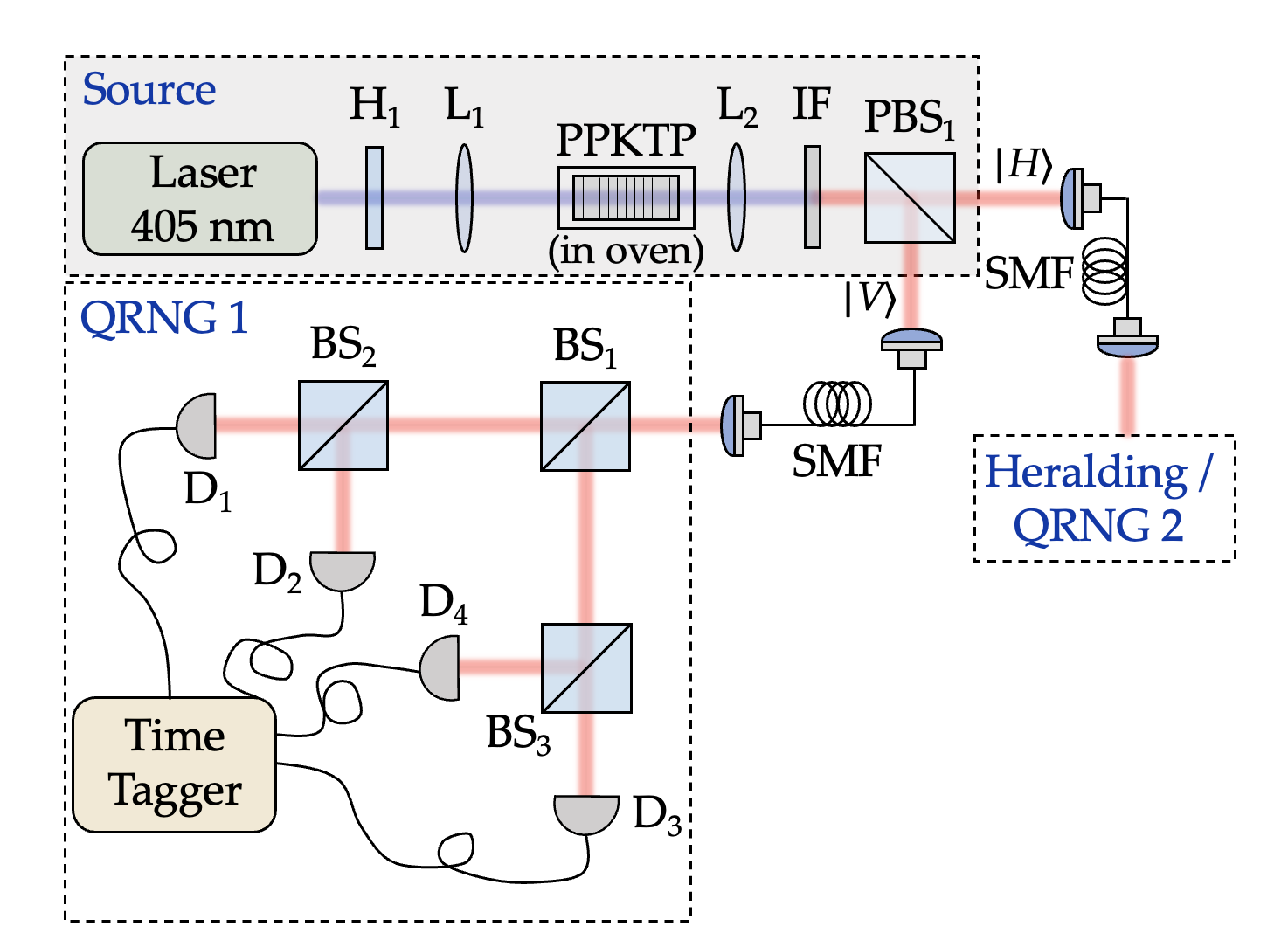}
    \caption{Schematic of the experimental setup.  {\bf Single photon source:} PPKTP nonlinear crystal enclosed in an oven is pumped by 405 nm laser. A half wave plate (H\textsubscript{1}) is used to control pump polarization and lens (L\textsubscript{1} of 300 mm) is used to focus the laser beam into the center of the crystal. Lens (L\textsubscript{2} of 35 mm) is used collimate the down-converted photons. A bandpass interference filter (IF of 810$\pm$10 nm) is used for filtering the pump. Polarization beam splitter (PBS\textsubscript{1}) is used to separate generated orthogonally polarized photons. {\bf QRNG unit}: $|V\rangle$ polarized single photons are coupled into single-mode optical fiber (SMF). Three non-polarizing 50:50 beam splitters (BS\textsubscript{1}-BS\textsubscript{3}) are used to generate two-bit random numbers. Four single photon counting modules (D\textsubscript{1}-D\textsubscript{4}) are used to detect single photons at the output of BS\textsubscript{2} and BS\textsubscript{3}. The output of the SPCMs are fed to a time-correlated single photon counter (Time Tagger). \label{fig:Expt} }
\end{figure}
\begin{figure}[ht!]
   \centering
    \includegraphics[width=0.47\textwidth]{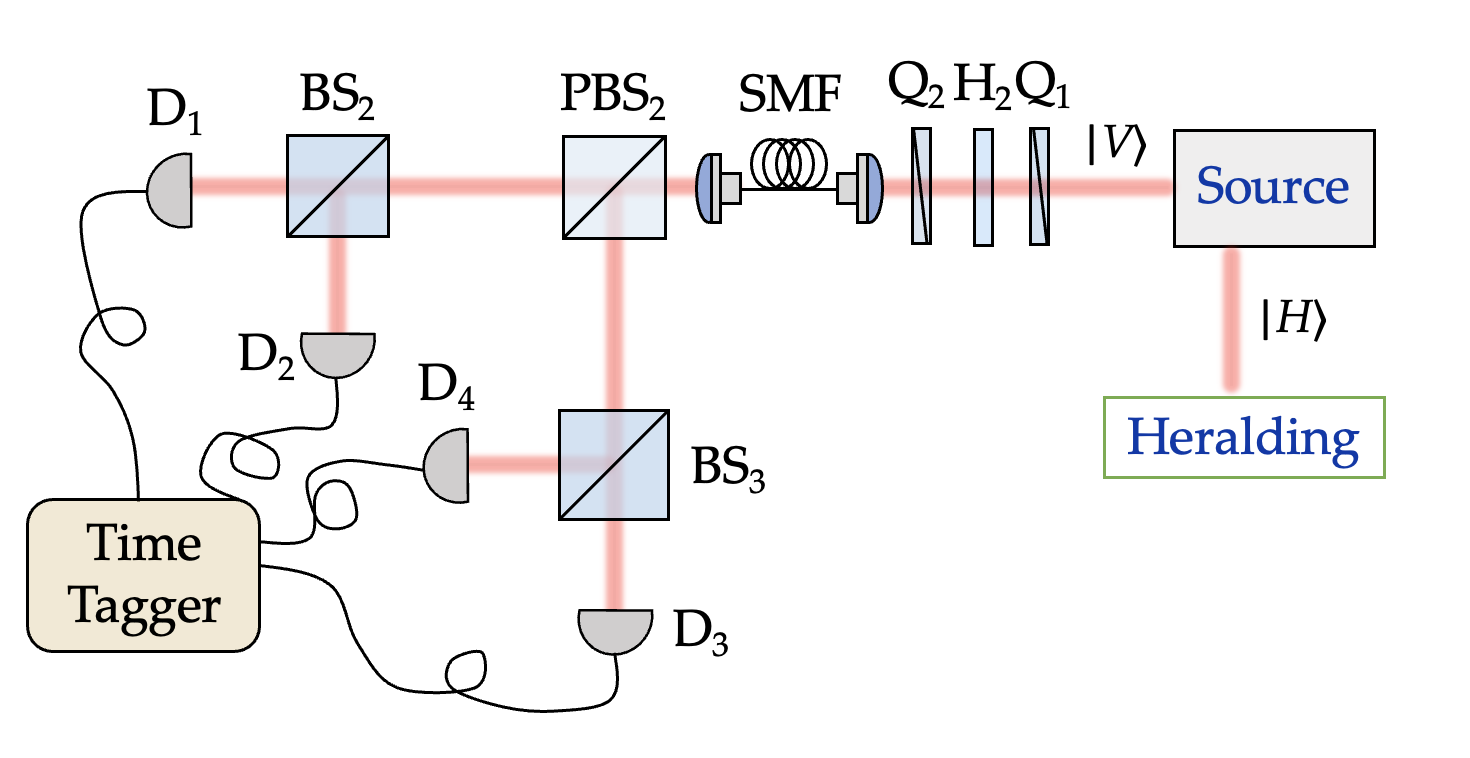}
    \caption{Schematic of the experimental setup to control the visibility of the path entangled state.  The whole QRNG unit is identical to the main setup except for the use of PBS in place of BS\textsubscript{1} and the combination of Q-H-Q is used to control the input polarization of the source. }
   \label{fig:ExptNoise}
\end{figure}
\section{Experiment}
\label{Expt}
\subsection{Setup}
We used single photons generated from SPDC process to experimentally demonstrate a path-entangled multi-bit quantum random number generator. The schematic of the experimental setup is shown in  Fig.\,\ref{fig:Expt}. A 10-mm-long PPKTP nonlinear crystal (Raicol) with an aperture size of 1\texttimes2 mm\textsuperscript{2} was pumped by a continuous-wave diode laser (Surelock, Coherent) at 405 nm. The crystal was type-II phase-matched with a poling period of $\Lambda = 10$~{\textmu}m. The pump laser was spatial-mode-filtered by coupling to a single-mode optical fiber (SMF) and a half-wave plate (H$_{1}$) was used to set the pump polarization. A plano-convex lens (L$_{1}$) of 300 mm was used to focus the laser into the center of the crystal with a spot size of 85~{\textmu}m. The variation in the wavelength of down-converted photons with temperature was studied using a spectrometer and the temperature was maintained at 39{\textdegree}C for obtaining down-converted photons at 810 nm. The orthogonally polarized photon pairs ($|H\rangle$, $|V\rangle$) generated were  collimated using a plano-convex lens (L\textsubscript{2}) of 35 mm and separated using a polarization beam splitter (PBS\textsubscript{1}). The $|V\rangle$ polarized single photons were coupled into a single-mode optical fiber (SMF) or multi-mode optical fiber (MMF) using appropriate collection optics. A bandpass interference filter (IF) at 810 nm center wavelength with a bandwidth of 10 nm FWHM was used for filtering the residual pump light. It should be noted that only $|V\rangle$ polarized single photons are used as a source to generate path-entangled photons. The $|H\rangle$ photon in our setup is used for heralding when heralded photons are used for QRNG and it is left unused when unheralded photons are used for QRNG. In such case, it can serve as a source for second independent path-entangled QRNG unit (QRNG 2). Maintaining fixed temperature or 810 nm of the down converted photons are not crucial for the experiment, a deviation of temperature by $\pm$15{\textdegree}C for PPKTP crystal did not alter much of our results. 

\begin{figure}[ht!]
	\centering
	\subfloat[][]{\label{fig:SMF} \includegraphics[width=0.48 \textwidth]{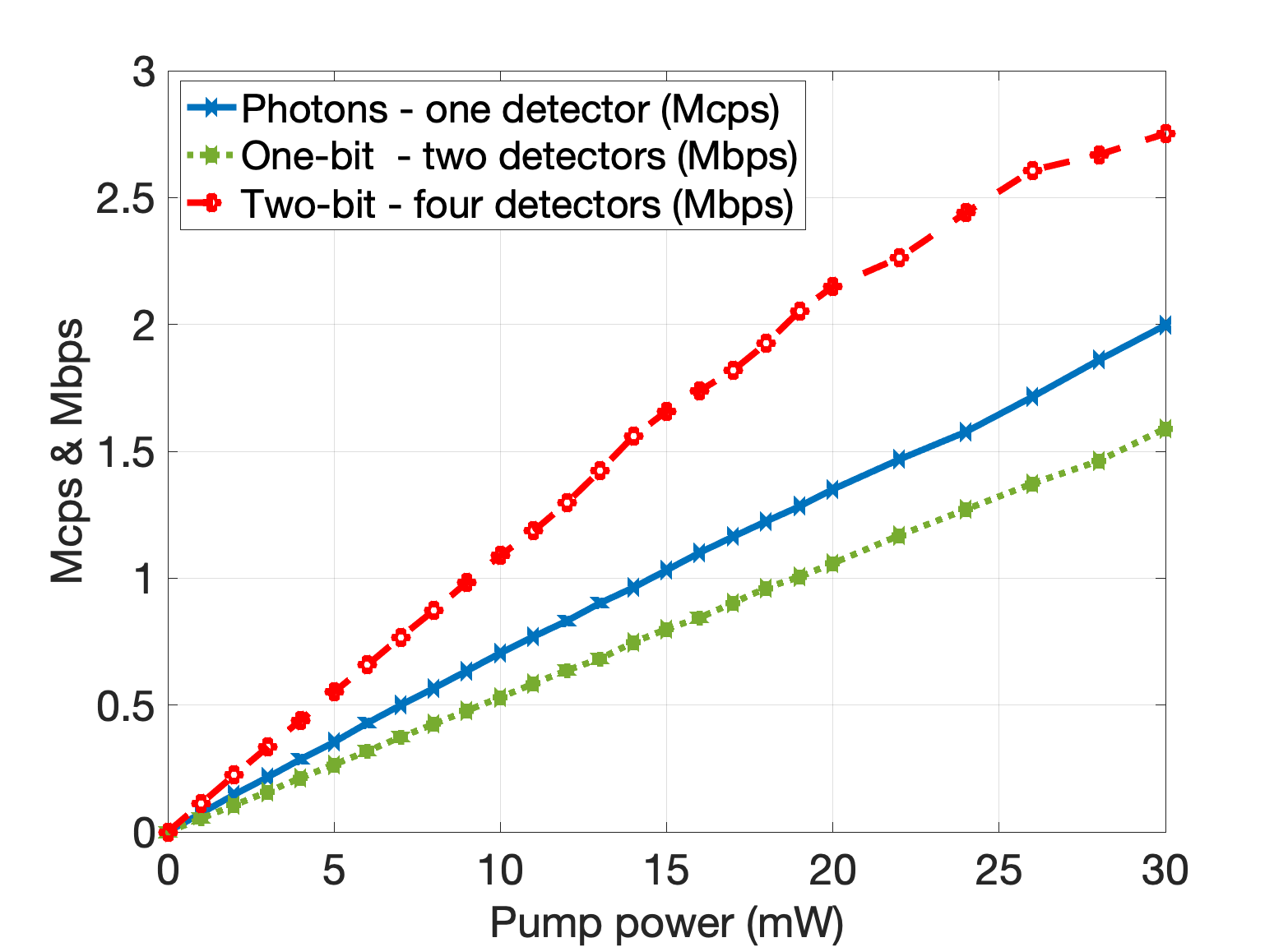}} \\
	\subfloat[][]{\label{fig:MMF} \includegraphics[width=0.48 \textwidth]{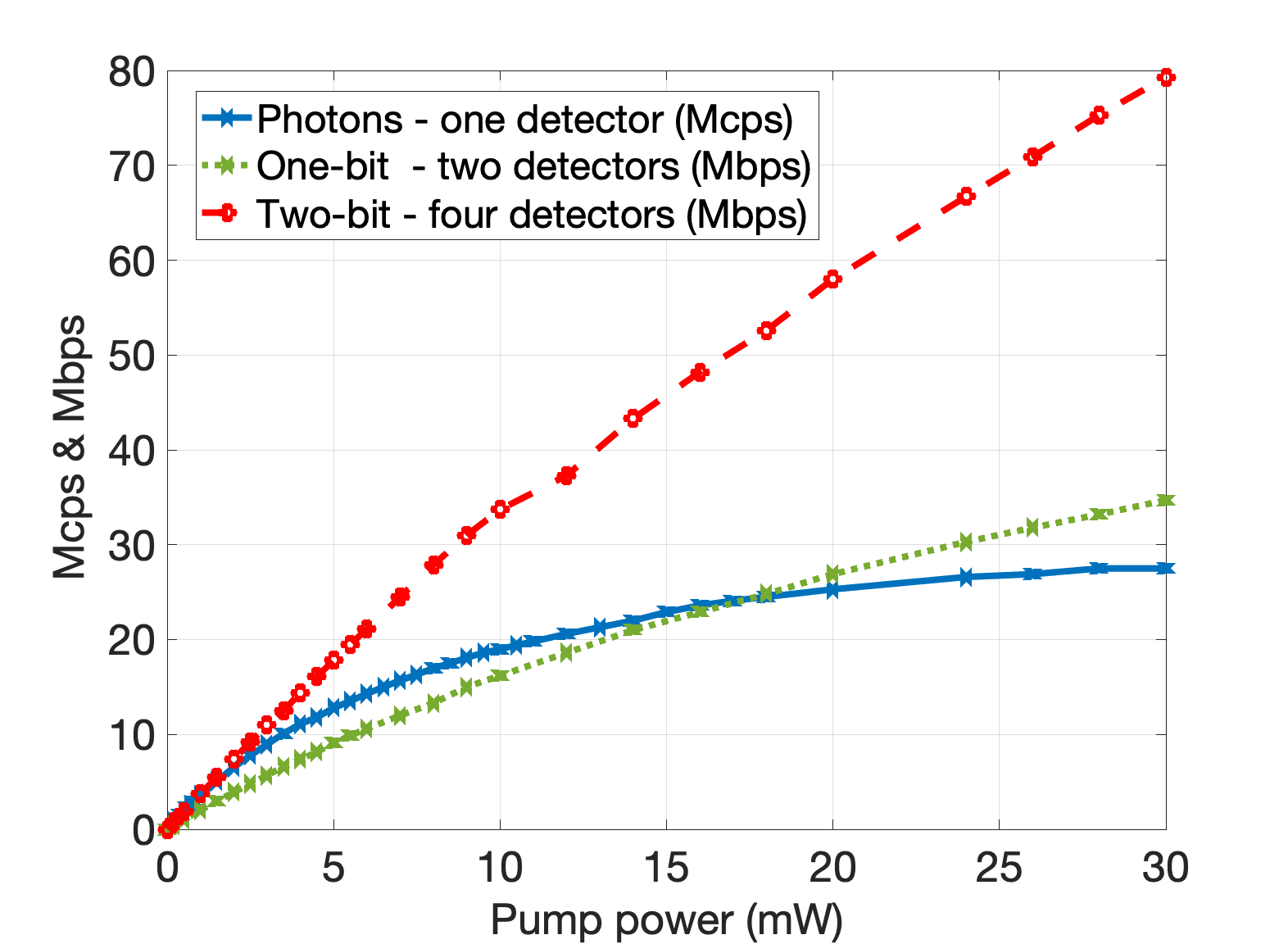}}
	\caption{This plot illustrates the number of photons recorded per second in one SPCM (blue solid), two SPCMs (green dotted) and four SPCMs (red dash-dotted) with increasing pump power. The photons are collected by coupling \protect\subref*{fig:SMF} SMF and \protect\subref*{fig:MMF} MMF to the QRNG source. The raw data for one-bit generation is possible from photon counts obtained through two SPCMs and two-bit generation requires four SPCMs. Due to the path-entanglement and two-bit mapping, one can see nearly 1.7 times higher bits generation compared to the photons detected when SMF is used and when counts were well below the detector saturation point. When MMF is used at a high power in this setup, we eliminate the limitation imposed by the dead time of each detector which would have caused in a loss of number of photon detections. Distributing the detection events across multiple paths increases the effective saturation point by spreading the photon among all detectors. Thus, we could record around a total of 80 Mbps, while the single photon detectors have a saturation point close to 28 Mcps.}
	\label{fig:SMF-MMF} 
\end{figure}

For a two-bit QRNG unit from single photon source, a configuration of three non-polarizing 50:50 beam splitters (BS\textsubscript{1}- BS\textsubscript{3}) in the path of single photons was used as shown in the Fig.\,\ref{fig:Expt}. Single photons at the output of the beam splitters BS\textsubscript{2} and BS\textsubscript{3} were coupled to multi-mode optical fibers (MMF) and detected using four single photon counting modules (SPCM-800-44-FC, Excelitas) with efficiencies of 65\%, dead time of 22 ns and dark counts {\textless} 100 Hz. The output of the SPCMs were fed to a time-correlated single photon counter (Time Tagger, Swabian instruments), and the arrival times of the photons were recorded with a time resolution of 4 ps. The same setup was used to generate one-bit random numbers by detecting photons at the output of the beam splitter BS\textsubscript{1}.

The effect of  noise on path-entangled state can be mimicked by controlling the photon detection visibility in the path-entangled photons.  A schematic of the experimental setup to control the visibility of path-entangled state is shown in Fig.\,\ref{fig:ExptNoise}.  The polarization degree of freedom is used to control the photon's superposition in path. In the main QRNG unit, BS\textsubscript{1} is replaced by the PBS, and a combination of Q-H-Q (two quarter-wave plates and a half-wave plate) are used to control the input state that can resulting in asymmetry in photons superposition along the paths.  By varying the angles in the wave plates, visibility of path-entangled state can be controlled. 

\begin{figure*}[!ht]
	\includegraphics[width=0.98 \textwidth]{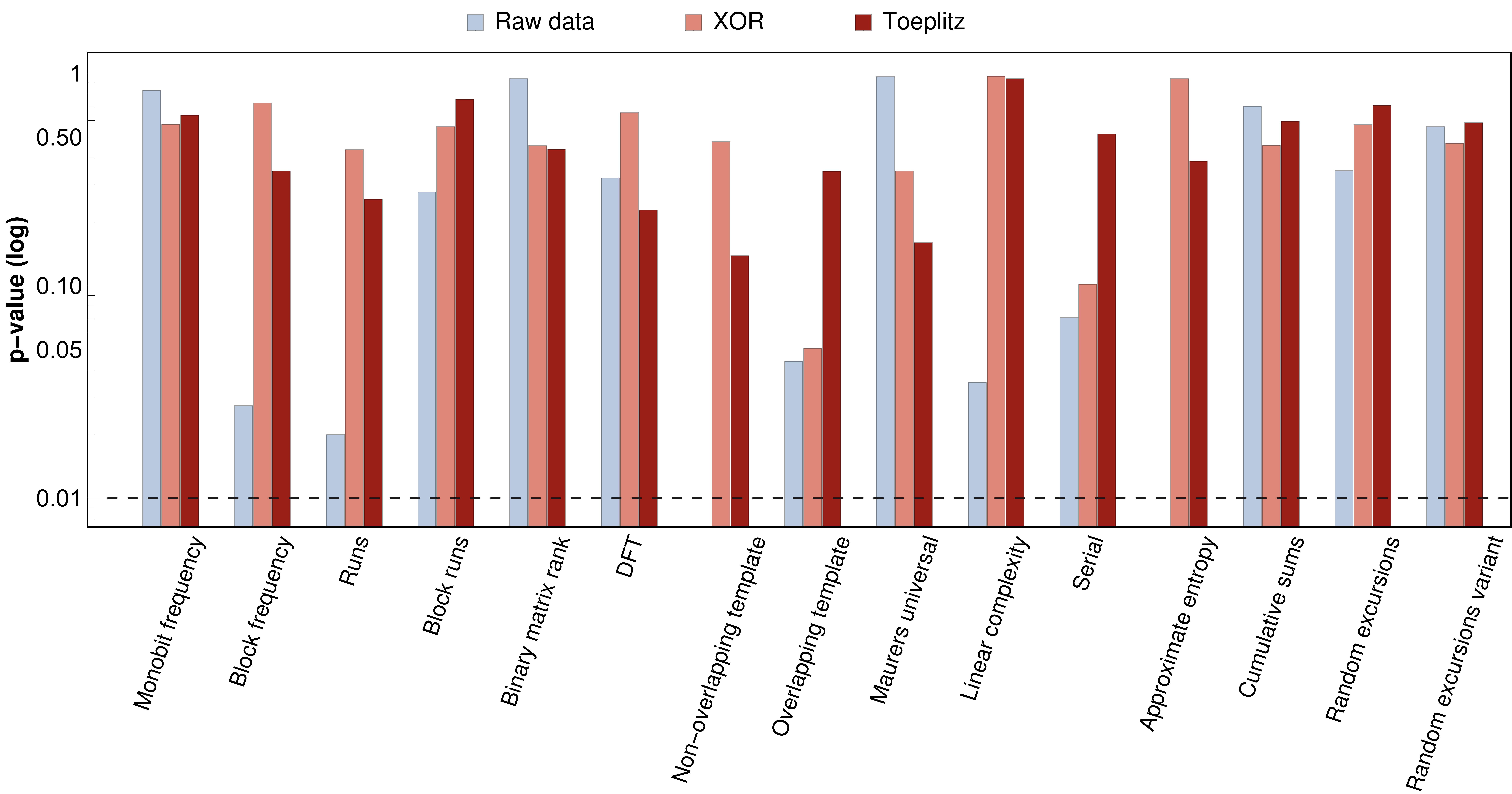}
	\caption{Review of NIST randomness test results for the raw and post-processed two-bit sequences generated at a rate of 1 Mbps recorded from the arrangement as shown in Fig.~\ref{fig:ExptNoise}. The raw data is of 3 Mbits long and passes 13 out of 15 tests. Toeplitz post-processing brings the bit length down by only $0.5\%$ (post-processed bit length $\sim$ 2.985 Mbits) while passing all the tests. XOR post-processing also passes all statistical tests but results in a loss of 50\% bits. The dashed line ($p =$ 0.01) represents the critical $p$-value above which the sequence is considered random. These results indicate that no statistical patterns were found in the tested samples of random bits. The same analysis done on two-bit generated at 50$\%$ visibility in the experimental setup for noise revealed that raw data passes only 5 tests while XOR post-processed data passes 11 tests. Similarly, the raw data generated at 60$\%$ visibility passes only 4 tests, but is able to pass 12 tests after XOR post-processing.}
	\label{fig:nistResults}
\end{figure*}

\subsection{Randomness tests on experimental data}

After the experiment was set up as illustrated in the preceding section, data was acquired for a wide range of controllable parameters in the setup. The photons originating from the PPKTP crystal were passed through the first beam-splitter (BS\textsubscript{1}) and the counts were recorded with time tagging along the paths from the second (BS\textsubscript{2}) and third (BS\textsubscript{3}) beam-splitters, which were then passed on for bit assignment and subsequent post-processing. As per the preset, bit assignment of 0 and 1 for each click on detectors D\textsubscript{1} and D\textsubscript{2} along the output of BS\textsubscript{1} (not shown explicitly in the schematic) were acquired to generate one-bit random numbers. Two-bit data was acquired with bit mapping of $00$, $01$, $10$ and $11$  for the clicks on the detectors D\textsubscript{1}, D\textsubscript{2}, D\textsubscript{3} and D\textsubscript{4} as shown in Fig.\,\ref{fig:Expt}. A series of statistical tests were carried out on various rates of bit-generation for prolonged time duration's to ascertain randomness in the acquired data.  Raw data of length averaging between 1 Mbits and 3 Mbits passed around 12 - 13 of the 15 tests in the NIST suite.  Raw data passed all the 15 tests after the post-processing as discussed in Sec.~\ref{NIST_pp}. When the visibility of path-entangled state was reduced to 50\% the raw data passed only around 5 tests and even after XOR post-processing the bits generated passed only about 11 tests. The same data subjected to either repeated post-processing or stringent rules for removing non-random data passed all the 15 tests even for data with low visibility at the cost of post-processing time and significant loss of bits. 

Photon detection rates are limited by the detectors' detection dead time. In the proposed scheme,  photons in path-entangled state and two-bit mapping for each detection can approximately generate bits at up to 8 times the maximum rate of photons each detector can resolve.  To validate this advantage, we present the experimental data in Fig.\,\ref{fig:SMF-MMF}, showing the number of unheralded single photon counts detected with increasing pump power using SPCM before sourcing it to QRNG unit, as well as the number of one- and two-bit generated after sourcing it to the QRNG unit using two and four SPCM, respectively.  In Fig.\,\ref{fig:SMF}, a plot of single photon count, one-bit and two-bit photon counts/sec  with increase in power is shown when SMF was used to couple the source with the QRNG unit.  Due to the high filtering of photons with SMF, the counts remained well within the detectors' saturation limit.  Even after accounting for the loss of photons due to detector efficiency,  the advantage of using path-entangled photons with two bit mapping with each path is seen by registering approximately 1.7 times the number of photons generated.

In Fig.\,\ref{fig:MMF} we present the same result when MMF is used to couple the source with the QRNG unit. Since MMF does not result in a huge loss of photons, we can see that the number of photons detected begins to show saturation when the pump power crosses 15 mW.  This is an indication of arrival of multiple photons within the detector's dead time resulting in null contributions to the counts.  When a photon is in path-entangled state, the load on the detectors gets shared among two detectors for one-bit generation and among four detectors for two-bit generation.  Thus, the distribution of photons among four paths increases the effective photon resolution and saturation counts by as many as 8 times in the case of two-bit. It can be observed from the Fig.~\ref{fig:SMF-MMF} that the usage of two detectors for generating one-bit sequences, the saturation of photon counts occurs at a higher pump power. Likewise, employing four detectors in order to produce two-bit sequences pushes the saturation point further. This enabled us to record about 80 Mbps when single photon detection saturation was seen around 28 Mcps. The bit rate of the setup was limited by the maximum power of the pump, and with a more powerful pump, this setup would be able to record around 150--200 Mbps before saturating the detectors.  However, in the following section we will see the decrease in CHSH parameter with increase in the photon counts/sec due to increase in multi-photon detection. That will set the bit rate for which we can see violation of CHSH inequality. 

The output at 80 Mbps from two bit mapping for each detected photon, which was generated using pump power of 30 mW, passed 12 of the NIST suite tests in raw form, and could pass all 15 tests after XOR post-processing. The visibility of the state was found to be close to 100\%. However, one-bit raw data with the same power passed a lesser number of tests. We attribute this to saturation happening at the detector affecting the visibility of the quantum state.

	\subsection{\label{NIST_pp}Post-processing of experimental data}

In the case where random numbers are used for cryptographic or security applications, the randomness of the output data from a RNG becomes an important property as it has strong correlations with the security of associated protocols. A good RNG has three properties -- the output random numbers must be unpredictable, uniformly distributed, and uncorrelated. The latter two are proved by using statistical tests on the output of the RNG, which in our case are the tests in the NIST test suite for RNGs. In order to prove unpredictability, one must prove that in the absence of knowledge of the input, the maximum probability of an adversary correctly guessing the next output bit from the RNG must be $0.5$. 

For device independent QRNGs, an important parameter is the guessing probability $p_g$, which is the  probability of the most probable outcome. This is related to the min-entropy of the system, defined as $H_{min}=-2\log_2({p_g})$. The min-entropy can be used to quantify randomness in the system in the presence of noise\,\cite{XFH13}. The quantity, however, is not conditioned on the noise itself. We thus uniformly distribute the final randomness by the use of a randomness extractor function, to obtain a sequence of bits that is completely random. By the Leftover hash lemma with side information, a higher min-entropy corresponds to a higher extractable randomness independent of the system \cite{DRM13}. We thus use min-entropy to determine the number of extractable bits from a sequence of $n$ bits (in this case, $n=256$). In the one-bit QRNG setup, we are able to see from the results that the distribution is $\epsilon < 2^{-50}$ close to a uniform distribution. We are able to achieve an average data rate of $8.348$ Mbps with a XOR extractor and $1.143$ Mbps with a Toeplitz extractor (with multiprocessing). The data loss in the XOR extraction is $50.00\%$, and  the data loss in the Toeplitz extraction is $7.71\%$. In case of the two-qubit setup, however, we observe that the XOR post-processing achieves an average bitrate of $42.97$ Mbps, while Toeplitz extraction is able to achieve $6.42$ Mbps. The higher bitrates may be attributed to efficient processing of larger file sizes generated by the system. The Toeplitz matrix used in the procedure may be constructed by using a private key as a seed. It is possible to use a pseudo-random number generator (PRNG) output for this purpose, however, to eliminate the possibility of the PRNG being compromised, we have set aside a $511$-bit sequence to function as the key, and the first $255+m$ bits of it are used as a private key (where $m$ is the length of the output bitstring, and is determined by the min-entropy of the system). It may be noted that due to the Leftover Hash Lemma, it is possible to reuse the seed, and we need not re-generate the private key.

	\subsection{Certification of the device
	\label{sec:cert}}
\subsubsection{CHSH inequality test for path-entanglement}

CHSH parameter $S$ for an entangled system can be calculated from combination of detection events obtained with different setting spanning over all basis states,
\begin{align}
S = \lvert E (\theta, \delta) - E(\theta, \delta^{\prime}) \rvert  + \lvert E (\theta^{\prime}, \delta) + E(\theta^{\prime}, \delta^{\prime}) \rvert.
\label{bellpara}
\end{align}
Violation of CHSH inequality ($S \leq 2$) is observed when  $S > 2$. For path-entangled state used for QRNG and described in Fig.\,\ref{fig:Expt}, $E(\theta, \delta)$ can be calculated from $P_{ij}$, probabilities of different basis states of the system for parameters $\theta$ and $\delta$,
\begin{align}
E(\theta, \delta) = P_{00}(\theta, \delta) + P_{11}(\theta, \delta) -P_{01}(\theta, \delta) - P_{10}(\theta, \delta).
\end{align}
Parameters $\theta$ and $\delta$ can be introduced by replacing standard beam splitter with a  variable beam splitters in operational form,
\begin{align}
B(\theta) = \begin{bmatrix}
	~\cos(\theta) & ~~i\sin(\theta) \\
	i\sin(\theta) & ~~\cos(\theta)
	\end{bmatrix}.
\end{align}	
This procedure is identical to the scheme where two detector combination was used in repeated measurement setting to obtain values for all four basis states\,\cite{SGP21}. Variable beam splitter in place of $BS_{1}$ will introduce $\theta$ and identically rotated variable beam splitters in place of $BS_{2}$ and $BS_{3}$ will introduce $\delta$. However, resetting $\theta$ to $\pi/4 +\theta$ on the first variable beam splitter and $\delta = \delta$ we will get the state $\frac{1}{\sqrt 2}[|00\rangle + |11\rangle]$ when $\theta = \delta = 0$.

\begin{figure}[!ht]
	
	\subfloat[][]{\label{fig:heraldedSval}\includegraphics[width=0.50\textwidth]{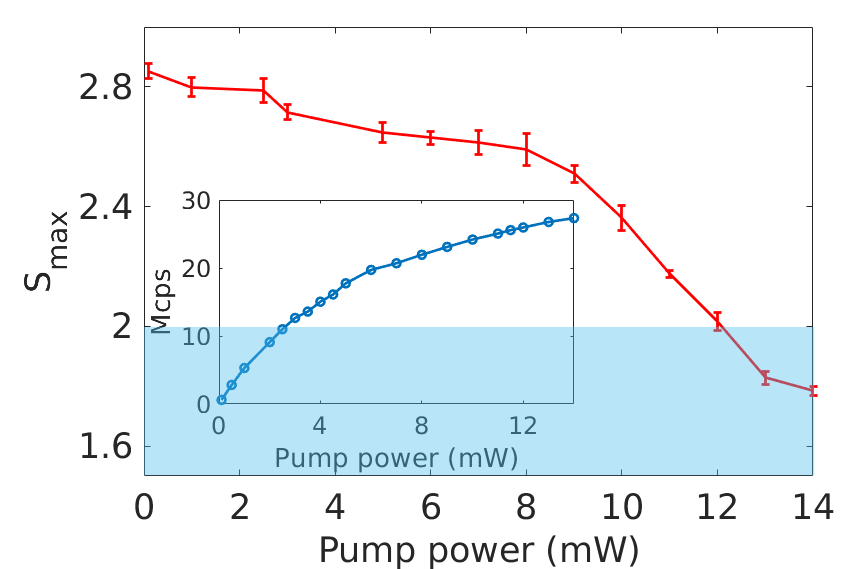}}
	
	\subfloat[][]{\label{fig:unheraldedSval}\includegraphics[width=0.50 \textwidth]{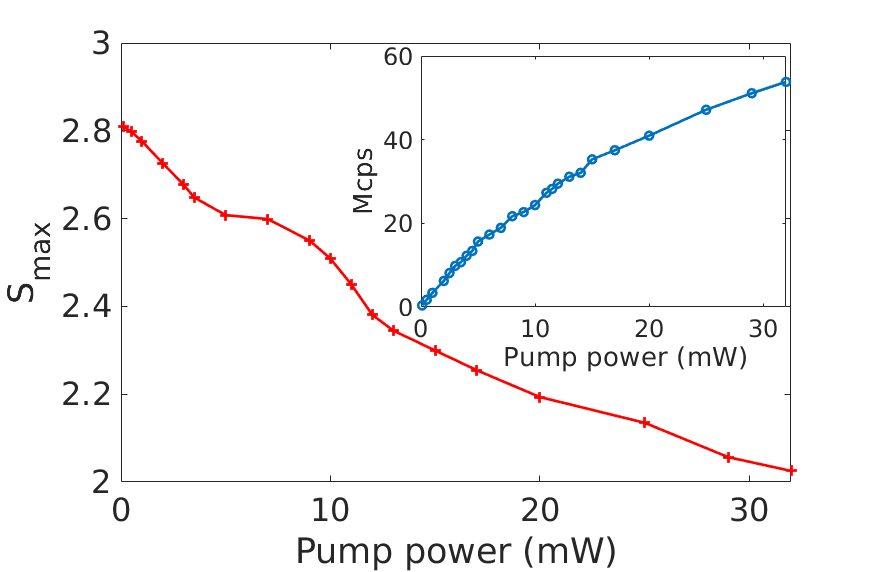}}
	
	\caption{Maximum value of CHSH parameter $S_{max}$ for \protect\subref*{fig:heraldedSval} heralded and \protect\subref*{fig:unheraldedSval} unheralded single photons in path-entangled state. The inset shows the photon counts recorded across 4 detectors for both the cases. With increase in pump power, $S_{max}$ is seen to decrease due to increase in detection of multi-photons across the detectors in the set time window (detector dead time). Due to saturation imposed by the heralding detector, a lower rate of bit generation is observed when compared to the case of using unheralded single photons as the source. The $S$ parameter ensures that the bit generation is limited to a rate that qualifies the entanglement certification.}
	\label{fig:puritySpara}
\end{figure}

In Fig.\,\ref{fig:heraldedSval} and \subref*{fig:unheraldedSval} we show the maximum value of CHSH parameter obtained experimentally with increase in the pump power (photon counts) for heralded and unheralded single photon in path-entangled state, respectively. In Fig.\,\ref{fig:heraldedSval} we note a decrease in CHSH parameter with increase in pump power. Initially the decrease is very small, the region where single photons are well separated in time and multi-photon events are very small. For higher pump power, multi photons detection in measurement window (detectors dead time) increases resulting in faster decrease in CHSH parameters and it goes below $2$ for pump power corresponding to photon detection rate of $\approx 27$ Mcps. This is also a saturation point of the single photon detector used for heralding. Since each detection corresponds to two-bit of random numbers, we can generate a maximum of $54$ Mbps of entanglement certified random numbers in our setup using heralded single photons in path-entangled state. When single photons are not heralded, we can bypass the detection limit of the heralding detector and generate more number of bits taking advantage of spread of photon detection along multiple paths. In Fig.\, \ref{fig:unheraldedSval} we show that photon detection rate of approximately $55$ Mcps are recorded when we still get $CHSH >2$. This corresponds to random number generation at the rate of $110$ Mbps. The CHSH parameter ensures that even when all the four detectors have not reached saturation point when unheralded photons are used, however the CHSH parameter decreases for higher pump powers due to an increase in detection of multi-photons.

\begin{figure}[!ht]
	\includegraphics[width=0.50 \textwidth]{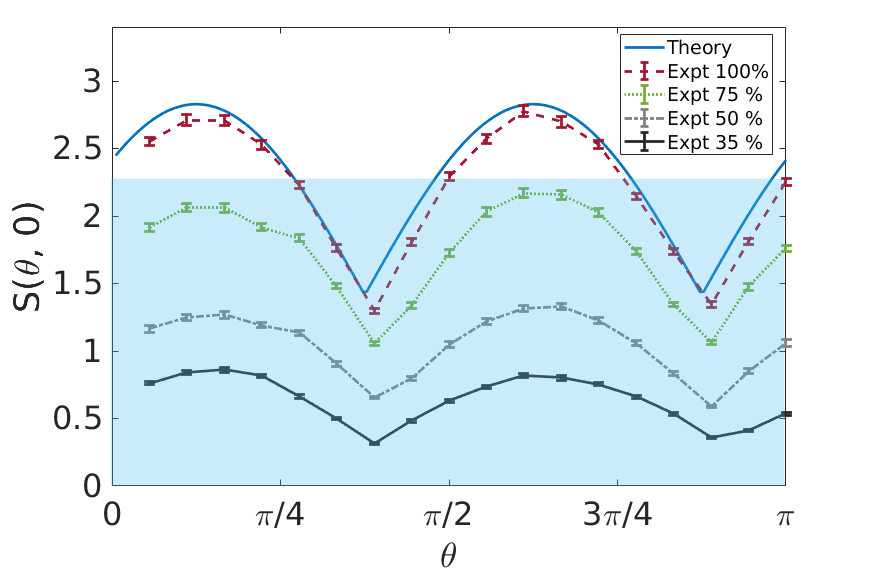}
	\caption{CHSH parameter S as a function of $\theta$ when $\delta = 0$, $\delta^{\prime} = \pi/4$ and $\theta^{\prime} = 3\pi/8$. Maximum violation of CHSH inequality is obtained when pairs $\theta$, $\theta^{\prime}$ and $\delta$, $\delta^{\prime}$ are orthogonal to each other. With decrease in entanglement visibility we see decrease in the maximum value of $S$ and transition from violation of CHSH inequality to validation of CHSH inequality is seen around when visibility crosses $70\%$.}
	\label{fig:puritySparaA}
\end{figure}

\begin{figure}[ht!]
	\centering
	\subfloat[][]{\label{fig:100pct} \includegraphics[width=0.46\textwidth]{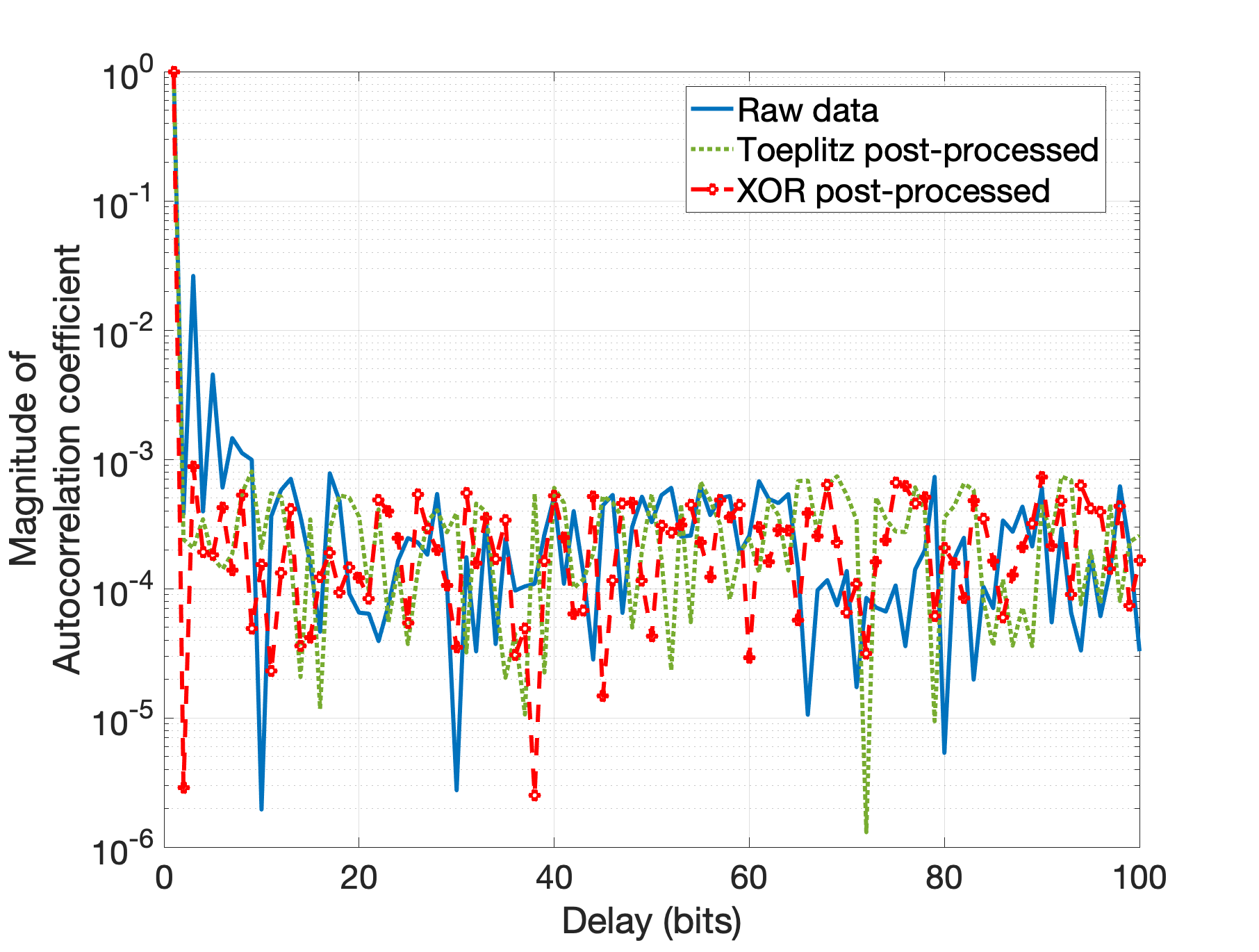}} \\
	\subfloat[][]{\label{fig:90pct} \includegraphics[width=0.46\textwidth]{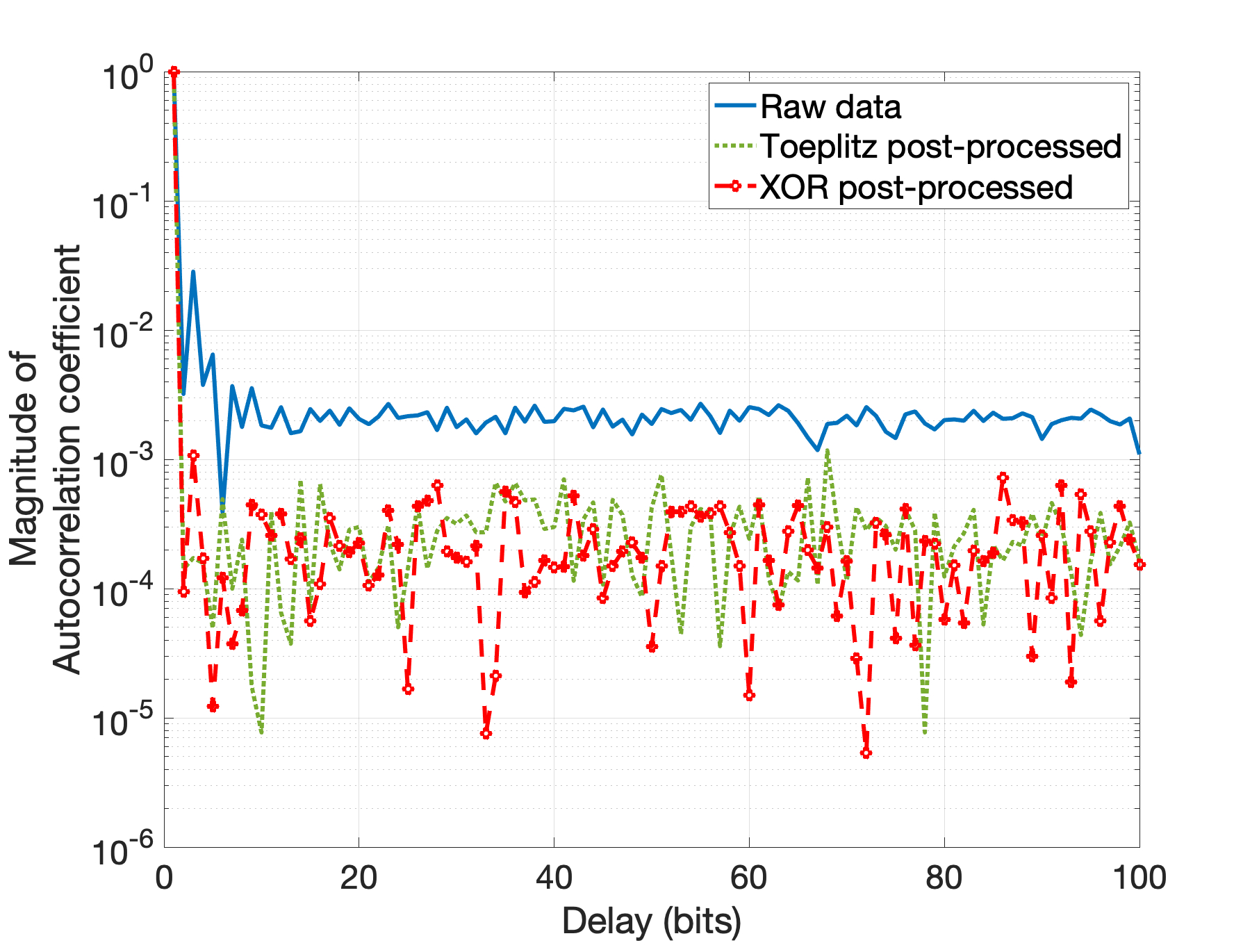}} \\
	\subfloat[][]{\label{fig:50pct} \includegraphics[width=0.46\textwidth]{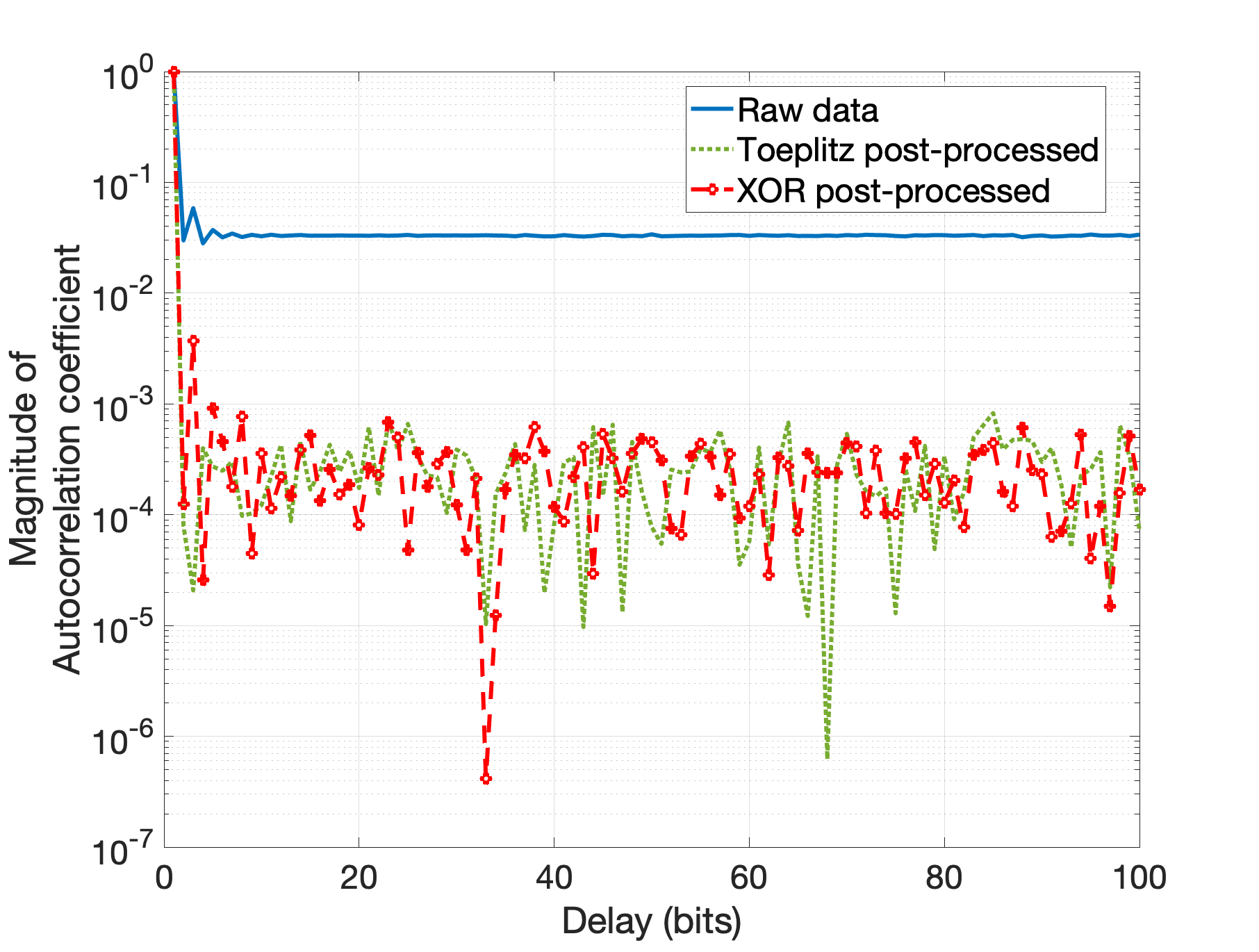}}
	\caption{Magnitude of autocorrelation coefficients corresponding to datasets generated at \protect\subref*{fig:100pct} 100\%, \protect\subref*{fig:90pct} 90\% and, \protect\subref*{fig:50pct} 50\% visibility, respectively. Autocorrelation between bits is higher for data obtained from lower visibility setups indicating suboptimal performance.}
	\label{fig:autocorrs_figs}
\end{figure}

In Fig.\,\ref{fig:puritySparaA} we show the theoretical expected value and experimental value for $S(\theta, 0)$ when $\theta^{\prime} = 3\pi/8$ and $\delta^{\prime} = \pi/4$. When $\theta = \pi/8$, that is when $\theta$ and $\delta$ are orthogonal to $\theta^{\prime}$ and $\delta^{\prime}$  we see maximum violation, $S = 2\sqrt{2} > 2$. For the above configuration $\delta =0$ and $\delta^{\prime}=\pi/4$ will be equivalent to absence of $BS_2$ and $BS_1$ and presence of $BS_2$ and $BS_1$, respectively. Parameter $S$ calculated from experimental output (Expt $100\%$) is in close agreement with theoretical value. Purity of path entangled state can be reduced by reducing the entanglement visibility using waveplates as shown in Fig.\,\ref{fig:ExptNoise}.
With decrease in entanglement visibility we see the decrease in the maximum value of $S$. Transition from  violation of CHSH inequality to validation of inequality is seen when visibility decreases below $70\%$, in agreement with previous results\,\cite{SGP21}. Fidelity of single photon events in attenuated laser source is lower than this threshold to show CHSH violation. 
This effectively self-certifies the device for the production of random bitstrings  based on measurement of a coherent entangled state.  As a result, the CHSH parameter $S$ serves as a reliable witness to the quantumness of the path-entangled QRNG setup. The threshold value of $S$ can thus be used as a certification that the randomness of the output contains a significant contribution from the quantum measurement. A value of $S$ greater than the classically expected maximum also provides a qualitative measure of entanglement, and therefore, purity of the state.  
	
	\subsubsection{Autocorrelation test}
Another metric to analyse the randomness of raw data is its autocorrelation. Fig.~\ref{fig:autocorrs_figs} shows a plot of the magnitude of  autocorrelation coefficients up to a delay of 100 bits for a subsequence formed from the first $10^7$ bits of the output sequence.  We find that beyond a delay of 100 bits, the mean and standard  deviation of the autocorrelation coefficients do not change  significantly. For a truly random sequence, the autocorrelation coefficients should have zero mean and a standard deviation of $1/\sqrt{N}$. For a $10^7$-bit sequence, this value is about $3.162\times 10^{-4}$. For raw device output, in the cases illustrated in Figs.~\protect\ref{fig:100pct}-\protect\ref{fig:50pct}, the standard deviation is found to be $2.684\times 10^{-3}$, $3.586\times 10^{-3}$, and $3.340\times 10^{-2}$ respectively.

It can be clearly observed from Fig.~\protect\ref{fig:100pct}-\protect\ref{fig:90pct} that the autocorrelation in the raw output is prominently seen for a delay of 2 bits  for the 90\% visibility case, which is not present in the 100\% visibility scenario. The case shown in \protect\ref{fig:50pct} corresponds to a suboptimal operational regime where the QRNG is unable to self-certify, as indicated by the high autocorrelation in the raw output from the device.  It can be observed from the figure that the QRNG when operating at high visibility (quantum regime) produces an output which is much closer to the extracted randomness than the output at lower visibility states (classical regime). QRNG output obtained at lower visibility is not able to pass all the NIST tests, even after XOR randomness extraction. Toeplitz extraction is still able to recover the randomness from this data, however, there are significant data loss at this point, and the QRNG functions sub-optimally.

We have measured the autocorrelation between the bit commitment assigned to the detector output. If multiple detectors were triggered simultaneously due to muti-photon events, then multiple bit-commitments would be made together, and one would be able to detect region of high autocorrelation at longer delays. In our data, we do not see significant variation in autocorrelation beyond a delay of 100 bits under full visibility. Therefore results from the raw bits rules out multi-photon events and protects QRNG from photon splitting attack.

\section{Conclusion}
\label{conc}

We have proposed a scheme for multi-bit QRNG from a controlled path-entangled single photon state.  An experimental demonstration of one-bit and two-bit QRNG from proposed scheme has also been reported.  In addition to being certified for quantumness via CHSH inequality violation,  the raw data generated from various different settings in the system have been able to pass about 12 of the 15 NIST suite tests used to certify randomness.   Using two different methods of post-processing, the generated random numbers are able to pass these statistical tests.  We also show a reduction in the number of tests passing with a corresponding decrease in entanglement visibility in the path-entangled  state.  Unlike other entanglement-based QRNG schemes that rely on coincidence events, path-entangled demonstration relies on single photon detection events resulting in higher bitrates of true random number generation. Furthermore, since the photons are produced by the SPDC process, the system can self-certify even in the presence of a poisoned source, making it much less vulnerable to quantum non-demolition measurement-based attacks as well as photon-number splitting attacks. The generation of single photons using an SPDC process is experimentally tunable, and thus one can control the output bitrate up to the maximum power of the pump laser or up to the detectors' saturation limit. Therefore, it is plausible for one to engineer the scheme as per their requirements without compromising on the quantumness in the generated states of photons.
 
The use of path-entangled states distributes the photon among the detectors, which allows our setup to bypass the bitrate limitation imposed due to the dead time of the detectors to some extent. This contributes significantly towards the generation of higher bit rates. In our results, we show the limit of photon detection at $\sim$ 28 Mcps in our detector is bypassed to generate $\sim$ 80 Mbps in path-entangled state. It is also possible to extend this scheme by setting up two independent two-bit QRNG units along each path of SPDC photons, making it a highly resourceful way to develop two completely independent QRNG units from a single SPDC photon source. On-chip tunable photon sources can be generated using directional couplers and nonlinear waveguides \cite{SST16}, which lends our scheme a lot of flexibility in terms of implementation platforms and can be extended to any other form of single photon source. It also enables one to consider the possibility of future on-chip integration with dedicated hardware-based post-processing.   

\begin{acknowledgments}
We acknowledge the support from the Office of Principal Scientific Advisor to Government of India, project no. Prn.SA/QSim/2020.
\end{acknowledgments}

\section*{Availability of data and materials}
Data sharing is not applicable to this article as no datasets were generated or analysed during the current study.

\section*{Competing interests}
The authors declare that they have no competing interests.


\end{document}